\documentclass[sigconf,nonacm]{acmart}

\usepackage{graphicx}
\usepackage{xcolor}
\usepackage{algorithm}
\usepackage{booktabs}
\usepackage{algorithmic}
\usepackage{subcaption,booktabs}
\usepackage{xurl}
\usepackage{float}
\usepackage{placeins}
\usepackage{balance}
\usepackage{enumitem}
\usepackage{hyperref}
\usepackage{enumitem}

\begin{document}
\title{Tempest: A GPU-Accelerated Engine for Streaming Temporal Random Walks}

\author{Md Ashfaq Salehin}
\orcid{0009-0000-7441-7798}
\affiliation{
    \institution{University of Sussex}
    \country{United Kingdom}
}

\author{George Parisis}
\affiliation{
    \institution{University of Sussex}
    \country{United Kingdom}
}

\author{Luc Berthouze}
\affiliation{
    \institution{University of Sussex}
    \country{United Kingdom}
}

\renewcommand{\shortauthors}{Salehin et al.}

\begin{abstract}
Temporal random walks, which sample causality-preserving paths, are widely used to analyze time-stamped interactions in domains such as microservices, finance, and online platforms. Generating such walks at scale is challenging because real-world graphs evolve as high-volume streams, making continuous ingestion, efficient memory usage, and strict temporal ordering essential for practical deployment. We present Tempest (TEMPoral nEtwork Streaming Traversals), a GPU-accelerated engine for streaming temporal random walks. Tempest combines a GPU-native dual-index organization over a shared edge store with a hierarchical cooperative scheduler that dispatches walks at thread, warp, or block granularity based on per-step node convergence, enabling efficient start-edge selection, hop-by-hop causality enforcement, and window-based eviction without synchronization. It further provides closed-form constant-time samplers for common temporal bias functions. Our evaluation demonstrates sustained real-time processing of billion-edge streams under sliding windows, outperforming prior systems in ingestion and walk generation throughput while preserving causal correctness.
\end{abstract}



\maketitle

\section{Introduction}
\label{sec:introduction}

Temporal networks are graphs in which interactions (edges) between nodes are time-dependent and occur at specific moments \cite{Holme12}. Unlike static networks where edges are assumed to persist, temporal networks explicitly encode the timing and order of interactions, capturing the dynamic nature of evolving systems. These networks can be either \emph{directed}, where interactions have a defined source and target (e.g., $u \rightarrow v$), such as in email exchanges, phone calls, transportation routes, financial transactions, and microservice calls; or \emph{undirected}, where interactions are mutual and symmetric, as in face-to-face social encounters or co-location events \cite{Holme12}.

Because temporal networks encode fine-grained time order, paths must preserve temporal causality. Reasoning over all such causality-preserving paths quickly becomes infeasible: even counting them is NP-complete, making exact enumeration impractical for large graphs \cite{enright2025counting, zhang2022reachability}. This intractability motivates the use of \emph{temporal random walks}, which provide a scalable alternative by sampling representative time-respecting paths without requiring full enumeration. Sampling not only keeps computation feasible by restricting to a fixed number of walks, but also provides diversity in exploration, ensuring that different regions and temporal patterns of the network are covered. These walks can be generated either in an \emph{unbiased} manner, where all valid temporal edges are chosen uniformly, or with \emph{temporal biases} that favor more recent interactions. Common examples include \emph{linear} and \emph{exponential} weighting \cite{ctdne2018}, while extensions such as \emph{temporal node2vec} incorporate additional structural return and in-out parameters \cite{temporal_node2vec2020}.

Temporal random walks are not only of theoretical interest but also form the basis of several temporal node embedding methods such as CTDNE and others \cite{ctdne2018, dynnode2vec2018, streamwalk2019,caw2021, neurtws2023}. These embeddings enable a variety of downstream tasks across different domains. For example, in microservice systems they support incident detection, root-cause localization, and workload forecasting \cite{luo2022workload, luo2024workload, dutt2025guide, somashekar2024gamma}, while in finance they have been applied to stock price forecasting \cite{xu2021rest}. As a result, the efficiency of temporal walk sampling is critical not only for embedding methods themselves but also for the scalability of these downstream applications.

Temporal random walks are also expensive to generate. Walk sampling can dominate training pipelines in large-scale graph learning systems, accounting for up to 96.2\% of end-to-end runtime \cite{bingo2025,gsampler2023} and up to 98.8\% of execution time in distributed implementations \cite{knightking2019, csaw2020}. A recent system reported that random walk computation alone required 3.5 hours, representing 35\% of overall processing time on graphs with hundreds of millions of nodes and billions of edges \cite{flowwalker2024}. The severity of this bottleneck has driven some temporal GNNs to bypass explicit walk sampling in favor of edge streaming or message passing \cite{tgn2020, hamilton2017inductive, xu2020inductive}.

\paragraph{Related work:}
Several systems accelerate random walk generation but target different problem settings. Static walk engines such as KnightKing~\cite{knightking2019}, GraphWalker~\cite{graphwalker2020}, FlashMob~\cite{flashmob2021}, and ThunderRW~\cite{thunderrw2021}, and GPU-based systems including CSAW~\cite{csaw2020}, Skywalker~\cite{skywalker2024}, and gSampler~\cite{gsampler2023}, achieve high throughput on static graphs but do not account for temporality. Dynamic graph engines such as FlowWalker~\cite{flowwalker2024} and Bingo~\cite{bingo2025} support evolving structures but do not take timestamps as input and cannot generate causality-preserving walks. The only published systems supporting causality-preserving temporal random walks at large scale are TEA~\cite{tea2023} and TEA+~\cite{teaplus2024}, which share the same engine; both are CPU-based and require loading the full dataset into memory.

To enable high-throughput walk generation, most existing walk engines rely on execution models in which the graph is first reorganized into a CSR-style representation to facilitate fast graph traversals during sampling, as is common in systems such as GraphWalker~\cite{graphwalker2020}, ThunderRW~\cite{thunderrw2021}, FlashMob~\cite{flashmob2021}, Skywalker~\cite{skywalker2024}, gSampler~\cite{gsampler2023}, and FlowWalker~\cite{flowwalker2024}. This preprocessing step allows efficient random access during execution but introduces a global reorganization cost performed ahead of time and typically excluded from reported metrics. As a result, these systems are primarily evaluated in batch settings and are not designed for streaming or real-time scenarios, where edges arrive continuously and walk generation must proceed without costly preprocessing.

In addition, the scale of real-world temporal graphs poses significant challenges for existing methods. Alibaba's 2022 microservice logs record 81 billion interactions over 14 days, representing only 0.5\% of total interactions during that period \cite{luo2022workload}. Similarly, Visa's payment infrastructure processed 234 billion transactions in 2024, averaging 639 million timestamped interactions per day across a continuously evolving network of credentials and merchant endpoints \cite{visareport2024}. Bulk-oriented methods that require loading the entire dataset into memory are not viable at this scale, making streaming ingestion with bounded memory and data eviction necessary.

These limitations motivate a practical temporal random walk system. First, it must strictly preserve temporal causality so that every sampled walk respects time order, unlike static walk engines that ignore temporal ordering. Second, it must sustain streaming-scale execution under bounded memory by efficiently ingesting new edges and evicting outdated ones; this setting is widely studied as \emph{sliding window processing} \cite{slidingtriangle2021, slidingtemporal2023, neurostream2023}. Finally, leveraging GPU-native designs is desirable for achieving high throughput at large scale, which in turn requires data structures and algorithms specifically tailored for GPUs to ensure that temporal ordering can be handled efficiently without degrading performance \cite{nvidia2012bestpractices}.

In this paper, we present Tempest (TEMPoral nEtwork Streaming Traversals)\footnote{Source code: \url{https://github.com/ashfaq1701/tempest}. Python package: \url{https://pypi.org/project/tempest-rw/} (\texttt{pip install tempest-rw}).}, a GPU-accelerated system for temporal random walk sampling. Our contributions include:

\begin{itemize}[leftmargin=*]

\item \textbf{A GPU-native temporal random walk engine} built on a dual-index edge store with two views over a shared edge array.

\item \textbf{Hierarchical cooperative scheduling}, an execution model that groups walks at the same node and step into a unit of work and dispatches that unit at thread, warp, or block granularity.

\item \textbf{A bounded-memory streaming architecture} that integrates batch ingestion, window eviction, and walk generation under sliding-window semantics, materializing GPU-efficient index layouts at each window update without fine-grained mutation or coordination overhead.

\item \textbf{Comprehensive experimental validation} demonstrating scalability to datasets with up to 81 billion temporal edges, more than sixty times larger than those used in prior work, together with detailed ablation study and sensitivity analyses that explain performance trade-offs.

\item \textbf{Open-source implementation} released as a Python library with a high-performance C++ core and comprehensive APIs.
\end{itemize}

\section{Tempest: System Architecture}
\label{sec:architecture}
This section presents the system architecture of Tempest. We first formalize the execution constraints imposed by causality-preserving temporal random walks on GPUs, then describe the dual-index data structure over which walks operate. We then present the walk scheduler, which dispatches walks across cooperative GPU execution tiers based on per-step node convergence. The remainder of the section covers temporal bias sampling, streaming ingestion under sliding-window semantics, and a complexity analysis of walk execution, index construction, and memory usage.

\subsection{Design Constraints of Temporal Walk Execution}
\label{sec:design-constraints}

We consider a temporal graph
\[
G = (V, E_T), \qquad E_T \subseteq V \times V \times \mathbb{R}^+ ,
\]
where each temporal edge $e = (u, v, t)$ represents an interaction from node $u$ to node $v$ occurring at time $t$.

For a temporal random walk that arrives at node $v$ at time $t$, causality requires that transitions respect temporal order. The causality-preserving neighborhood in the forward direction is
\[
\Gamma_t(v) = \{ (v, w, t') \in E_T \mid t' > t \},
\]
with the backward case defined analogously. Each hop selects its next edge from this node-conditioned and time-filtered neighborhood; ties on timestamps are broken by uniform random sampling.

The set $\Gamma_t(v)$ is hop-dependent: its membership depends on both the current node and the timestamp reached at the previous step. It cannot be precomputed, cached, or reused across hops without violating temporal correctness. On GPUs, enforcing hop-dependent temporal constraints by enumerating valid neighbors or synchronizing threads does not scale. These approaches produce irregular memory access and coordination overhead, which limit throughput on massively parallel hardware~\cite{nvidia2012bestpractices}.

\subsection{System Overview}
\label{sec:system-overview}

Tempest ingests a temporal edge stream $(u, v, t)$ and processes it in GPU-resident batches. Each batch undergoes three stages: GPU-parallel ingestion with sliding-window eviction (Section~\ref{sec:streaming-ingestion}), reconstruction of a dual-index representation over the active edge set (Section~\ref{sec:dual-index}), and generation of temporal random walks under a hierarchical cooperative scheduler (Section~\ref{sec:walk-scheduling}). The dual index provides both global temporal access (for start-edge selection and window eviction) and node-conditioned temporal access (for walk progression). Once built, the index is reused across many walk sampling operations, amortizing construction (Section~\ref{sec:complexity}).

This design favors bulk reconstruction over fine-grained mutation. Tempest rebuilds its index at batch boundaries, in contrast to dynamic graph systems that incrementally update adjacency structures. Construction and window maintenance run entirely through GPU data-parallel primitives, avoiding synchronization and sequential updates. The resulting layout is compact, predictable, and aligned with sliding-window eviction.

\subsection{Dual-Index Organization}
\label{sec:dual-index}

\begin{figure}[h]
    \centering
    \includegraphics[width=\linewidth]{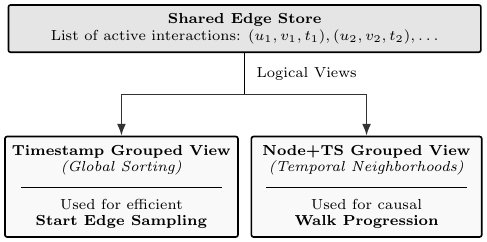}
    \caption{Dual-index organization. Two logical views over a shared edge store, distinguished by their offset arrays.}
    \label{fig:dual-index-overview}
\end{figure}

Tempest organizes active edges as a shared edge store with two logical views, illustrated in Figure~\ref{fig:dual-index-overview}: a \emph{timestamp-grouped view} and a \emph{node--timestamp-grouped view}. Both are offset structures over a single edge array; neither replicates edge data.

\paragraph{Timestamp-grouped view.}
The timestamp-grouped view orders edges globally by timestamp, with an offset array marking each timestamp group's boundary. Start-edge sampling under a temporal bias and bulk window eviction both proceed by binary search over the offset array: the bias selects a timestamp group, and the offset locates the corresponding edge slice in constant time. Window eviction then reduces to discarding the prefix of the edge array up to the temporal cutoff.

\begin{figure}[hbtp!]
    \centering
    \includegraphics[width=0.9\linewidth]{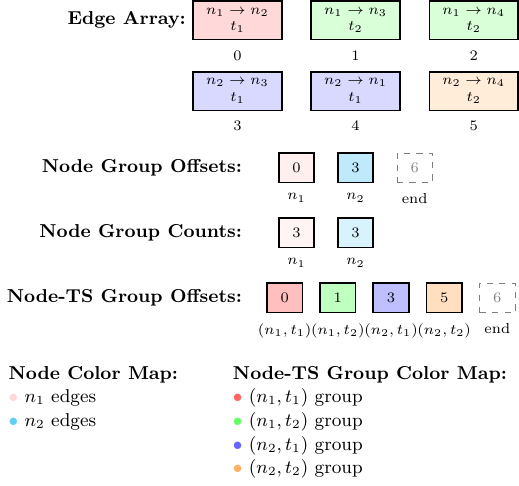}
    \caption{Node-and-timestamp-grouped view.}
    \label{fig:node-ts-grouped}
\end{figure}

\paragraph{Node-and-timestamp-grouped view.}
The node-and-timestamp-grouped view is a two-level offset structure. Edges are grouped by source node, then ordered by timestamp within each node's region. Figure~\ref{fig:node-ts-grouped} shows the layout. A node-group offset array locates each node's edge region; a node--timestamp-group offset array locates each timestamp sub-group within that region. Both offset arrays are built once per batch and reused across all walks.

A walk thread evaluates $\Gamma_t(v)$ in two steps. The node-group offset array gives the edge range $[a, b)$ for node $v$ in constant time. A binary search over the node--timestamp-group offset array within $[a, b)$ then locates the first timestamp strictly greater than $t$, returning $[c, b) \subseteq [a, b)$ as $\Gamma_t(v)$. The cost is $\mathcal{O}(\log G)$ where $G$ is the number of distinct timestamp groups at $v$. Selecting an edge from $\Gamma_t(v)$ is then constant-time pointer arithmetic. This two-stage lookup eliminates per-hop neighborhood scanning.

\subsection{Hierarchical Cooperative Scheduling}
\label{sec:walk-scheduling}

Causality-preserving temporal walks expose a tension when executed on a GPU. At each hop, a thread must locate $\Gamma_t(v)$ inside the node's edge range, sample under the active bias, and write the result. The work is short, dependent on the previous hop's timestamp, and concentrated on a single node's metadata. When tens of thousands of walks advance in parallel, threads in the same warp diverge across nodes, accesses to per-node metadata scatter across global memory, and control flow branches on walk lifetime. Coalesced access, shared-memory reuse, and uniform branching, all of which a GPU rewards~\cite{nvidia2012bestpractices, harrisreduction2007}, are absent. We introduce \emph{hierarchical cooperative scheduling}, an execution model that recovers them by treating walks at the same node and step as a unit of cooperative work and dispatching that unit at thread, warp, or block granularity based on how many walks have converged.

\begin{figure}[h]
    \centering
    \includegraphics[width=\linewidth]{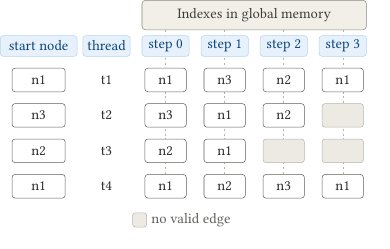}
    \caption{Full-walk execution where each thread advances one walk independently.}
    \label{fig:per-thread-walk}
\end{figure}

\subsubsection{The Full-Walk Baseline}
\label{sec:full-walk-baseline}

The naive GPU baseline is to assign one thread to each walk. We call this the full-walk baseline. At each step, the thread fetches its current node's edge range from global memory, locates the temporal cutoff $\Gamma_t(v)$ by binary search, draws an edge, and advances to the next hop. The kernel scales by launching more threads. Figure~\ref{fig:per-thread-walk} illustrates one step.

Walks that converge on the same node share the index range and the binary-search target, but full-walk threads fetch and search independently. On hub-skewed temporal graphs~\cite{Holme12, luo2022workload}, this redundancy dominates per-step global-memory traffic. Walks also terminate at different steps when $\Gamma_t(v)$ becomes empty, leaving lanes masked within an otherwise active warp.

\subsubsection{Cooperation and Its Challenges}
\label{sec:cooperation-and-challenges}

Walks that converge on the same node at the same step share work: the temporal binary search, the timestamp-group offsets it consults, and the edge-range arithmetic are identical across the group, while only the random draw and the picked edge differ per walk. NextDoor~\cite{nextdoor2021} exploits this on static graphs by caching the transit's adjacency list in shared memory across co-located walks. The temporal regime introduces five challenges that a static design does not face.

\begin{itemize}[leftmargin=*]
    \item \textbf{Variable walk length.} Temporal walks terminate early whenever no future valid edge exists at the current node, producing high per-walk variance in walk length. Under one-thread-per-walk dispatch, terminated lanes sit masked while their warpmates continue, so warp efficiency degrades step by step.

    \item \textbf{Runtime grouping.} The walk-population count $W$ at each node changes every step as walks migrate. It can only be obtained by inspecting walk state at the current step, so the grouping must be recomputed on every hop.

    \item \textbf{Granularity selection.} The execution unit that does the computation depends on $W$. A node with tens of walks underutilizes a block but suits a warp; a node with hundreds of walks overflows a warp's lanes. Suboptimal resource allocation reduces efficiency.

    \item \textbf{Bounded shared memory.} The per-node metadata that makes cooperation profitable is the timestamp-group offset array, whose size $G$ varies across nodes and across windows. Only some fit in a warp's slice of shared memory, fewer fit in a block's, and the rest must be served from global memory.

    \item \textbf{Mega-hubs.} A single node can hold thousands of walks. A one-block-per-task dispatch would monopolize an SM with one task while other SMs sit idle, requiring work to be split across blocks.
\end{itemize}

\subsubsection{Cooperative Dispatch}
\label{sec:cooperative-dispatch}

Tempest dispatches each per-node group of co-located walks to one of three execution units, sized to the GPU thread hierarchy: a single thread, a warp, or a thread block. The cooperative tiers (warp and block) preload the node's adjacency metadata into shared memory (smem) once per task; every thread of the unit reads from smem instead of global memory. With the metadata resident, the temporal binary search and the edge-range arithmetic become smem reads, and the heavy global-memory access is done once per node rather than once per walk.

\begin{figure}[h]
    \centering
    \includegraphics[width=0.95\linewidth]{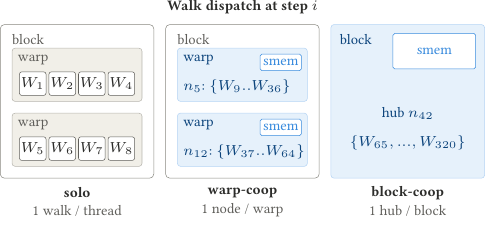}
    \caption{Three cooperative execution units. Solo dispatches one walk to one thread. Warp-coop assigns one node's walks to one warp, all 32 threads sharing the node's adjacency metadata in smem. Block-coop assigns one hub's walks to one block, with a larger smem allocation holding the metadata.}
    \label{fig:walk-dispatch}
\end{figure}

Figure~\ref{fig:walk-dispatch} shows the three units. Solo handles nodes carrying a handful of walks; the cooperation overhead is unjustified there, so each walk runs as in the full-walk baseline but launched within a step-bounded kernel. Warp-coop and block-coop differ in scale: a warp's smem allocation holds less metadata than a block's, so the warp tier serves moderate $W$ and the block tier serves the long tail.

\subsubsection{The Dispatch Plane}
\label{sec:dispatch-plane}

\begin{figure}[h]
    \centering
    \includegraphics[width=0.95\linewidth]{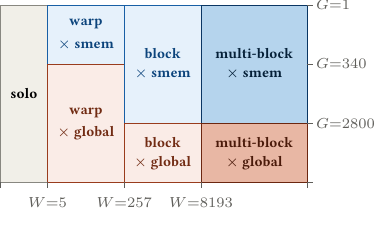}
    \caption{Dispatch plane. The $W$ axis selects the execution unit; the $G$ axis selects whether the node's adjacency metadata is preloaded into smem or read from global memory. W thresholds are obtained empirically (Section~\ref{sec:parameter-tuning}).}
    \label{fig:dispatch-plane}
\end{figure}

The choice between solo, warp-coop, and block-coop is governed by the walk-population count $W$. The choice between an in-smem metadata copy and a global-memory fallback within each cooperative tier is governed by the node's timestamp-group count $G$. The two together define a dispatch plane shown in Figure~\ref{fig:dispatch-plane}, partitioned into the five terminal kernels.

The $W$ thresholds set tier boundaries. The warp/block boundary is $W = \mathit{block\_dim}$: a node carrying more walks than a block can hold must use a block-tier kernel. The solo/warp boundary is a tunable hyperparameter that trades cooperation overhead against amortization gain at low $W$, where co-located walks are few. Both are tuned empirically (Section~\ref{sec:parameter-tuning}).

The $G$ thresholds set memory-tier boundaries. In the warp tier, the metadata sits in a warp's slice of the block's shared-memory allocation. In the block tier, it occupies the full block's allocation. Each is sized to the maximum that fits, giving the per-tier $G$ caps shown in Figure~\ref{fig:dispatch-plane}: above each cap, the metadata does not fit and the kernel reads from global memory instead. The block tier's larger budget tolerates roughly $8\times$ the $G$ that the warp tier does, reflecting the block hosting 8 warps that share the smem allocation.

Mega-hubs occupy the rightmost column. A node with $W > 8192$ walks would, under one-block-per-task dispatch, monopolize one streaming multiprocessor while others sit idle. Tempest splits such tasks into $\lceil W / 8192 \rceil$ disjoint sub-tasks, each dispatched to its own block. The walks at the node are partitioned into contiguous slices of up to 8192, with each block assigned a slice and performing sampling on it independently. The metadata is loaded once per sub-task because the underlying node is the same.

\begin{algorithm}[h]
\caption{Per-step walk scheduling.}
\label{alg:scheduler}
\begin{algorithmic}[1]
\REQUIRE walk array, step $i$, timestamp-group counts $G[\cdot]$
\ENSURE  task lists for the solo, warp-smem, warp-global, block-smem, block-global tiers

\STATE flag walks alive at step $i$
\STATE compact alive walks via \textsc{PartitionFlagged}
\STATE gather each alive walk's current node
\STATE \textsc{SortPairs} (current node, walk index) by node
\STATE \textsc{RunLengthEncode} on sorted nodes; \textsc{ExScan} on run lengths
\STATE partition runs by $W$ into solo, warp, block tiers \COMMENT{Fig.~\ref{fig:dispatch-plane}}
\STATE partition warp tier by $G$ into warp-smem, warp-global
\STATE partition block tier by $G$ into block-smem, block-global
\STATE expand block-tier tasks with $W > W_{\max}$ into disjoint sub-tasks
\end{algorithmic}
\end{algorithm}

\subsubsection{Per-Step Pipeline}
\label{sec:per-step-pipeline}

The scheduler produces the five terminal task lists with a sequence of GPU data-parallel passes, summarized in Algorithm~\ref{alg:scheduler}. Most stages are CUB primitives~\cite{cub}; the remaining are custom kernels (alive flagging, current-node gather, the partitions, and the mega-hub expansion). Two host syncs are required per step: one to read the alive count after compaction, and one to read the per-tier task counts before launching the terminal kernels.

\subsubsection{Comparison to Prior Dispatch Schemes}
\label{sec:comparison-prior-dispatch}

Multi-granularity dispatch on GPUs is itself not new. Skywalker~\cite{skywalker2024} and FlowWalker~\cite{flowwalker2024} assign work at warp or block granularity, but the dispatch input is the static structural degree of the current vertex and does not depend on the state of the walk population. Tempest's $W$ is a runtime, per-step quantity that requires inspecting walk state at the current step. The same node receives different dispatch granularities at different steps as walks migrate.

NextDoor~\cite{nextdoor2021} performs the closest analog of Tempest's per-step regrouping: it inverts the sample-to-transit map and dispatches each transit vertex to a warp, block, or grid of blocks based on the number of samples sharing that transit. Once a kernel choice is made, NextDoor caches the transit's adjacency slice unconditionally within that choice. Tempest extends the regrouping idea to the temporal regime: the data preloaded is the per-node timestamp-group offset array under sliding-window expiration, the decision to preload at all is governed by $G$, and Tempest filters walk state per step to drop completed walks before kernel launches.

FlexiWalker~\cite{flexiwalker2026} is recent work that adapts at runtime per node. The signal and the decision space differ. FlexiWalker selects between rejection and reservoir sampling per node using a runtime cost model, targeting dynamic random walks on static graphs. Tempest selects an execution unit and a memory tier per task based on $W$ and $G$, on streaming temporal graphs under sliding windows.

\subsection{Temporal Bias Sampling}
\label{sec:temporal-bias}

Each hop selects the next edge under a temporal bias evaluated within the index range exposed by the dual-index. Tempest's default sampler is \emph{weight-based}: it applies inverse transform sampling on the cumulative weight array of $\Gamma_t(v)$. Edges are assigned weights $w(e_i) = \exp(t_i - t_{\min})$ where $t_{\min} = \min_{e \in \Gamma_t(v)} t(e)$; the cumulative array $W[k] = \sum_{j=0}^{k} w(e_j)$ is built once per neighborhood, and a draw $u \sim \mathcal{U}(0,1)$ selects the smallest index $k$ with $W[k] \ge r$ for $r = u \cdot W[n-1]$ via binary search, at $\mathcal{O}(\log n)$ per hop.

When timestamp gaps within a neighborhood are uniform, only ordinal position matters, and the \emph{index-based} sampler admits closed-form inverse CDFs over the index $i \in [0, n)$:
\begin{equation}
\textbf{Uniform:} \quad i = \lfloor u \cdot n \rfloor .
\label{eq:index_uniform}
\end{equation}
\begin{equation}
\textbf{Linear:} \quad i = \left\lfloor \tfrac{-1 + \sqrt{1 + 4u n (n+1)}}{2} \right\rfloor .
\label{eq:index_linear}
\end{equation}
\begin{equation}
\textbf{Exponential:} \quad i \approx \lfloor n + \ln(u) - 1 \rfloor .
\label{eq:index_exponential}
\end{equation}
These collapse the per-hop cost to $\mathcal{O}(1)$ on a single random draw. Derivations and the numerically stable forms are deferred to the supplementary material.

Tempest supports Temporal Node2Vec~\cite{temporal_node2vec2020}, which adds a second-order bias $\beta(u, w)$ depending on the previous node. We apply $\beta$ via rejection sampling on the static exponential proposal: each hop accepts with probability $\beta(u, w) / \beta_{\max}$, where $\beta_{\max} = \max(1/p,\, 1,\, 1/q)$, keeping the inner CDF prev-independent so Node2Vec runs through the same cooperative dispatch path as other pickers.

\subsection{Streaming Ingestion and Window Management}
\label{sec:streaming-ingestion}

Tempest maintains an active temporal window of fixed duration $\Delta$. At time $t$, the active edge set is
\[
W(t) = \{ e \in E_T \mid t - \Delta \le t_e \le t \},
\]
where $t_e$ is the timestamp of edge $e$. Only edges within $W(t)$ participate in indexing and walk sampling, bounding memory usage by $|W(t)|$ regardless of total stream length.

Incoming batches advance the window forward in time. Each batch is sorted by timestamp before merge; edges older than $t - \Delta$ at merge time are dropped. The system assumes monotonic batch boundaries: edges arriving in a later batch with timestamps before the current window are treated as too late and dropped without retraction. This is consistent with the streaming-graph processing model used in prior work on sliding-window analytics~\cite{slidingtriangle2021, slidingtemporal2023}.

Each batch triggers a bulk reconstruction of the dual-index over the updated edge set rather than incremental mutation. Ingestion and rebuild costs depend on $|W(t)|$ and do not accumulate across batches; if per-batch processing time stays below the arrival interval, the stream sustains stable throughput without backlog (Section~\ref{sec:streaming_evaluation}). The window duration $\Delta$ trades temporal context against memory and rebuild cost; we measure this trade-off in Section~\ref{sec:window-size-sensitivity-analysis}.

\subsection{Complexity Analysis}
\label{sec:complexity}

\paragraph{Walk execution.}
Let $L$ denote the walk length and let $|\Gamma_t(v)|$ be the size of the causality-preserving neighborhood at a hop from node $v$. The per-hop cost depends on the sampler. The index-based picker performs a constant number of arithmetic operations and one random draw, yielding $\mathcal{O}(1)$ per hop and $\mathcal{O}(L)$ per walk. The weight-based picker performs a binary search over the cumulative weight array of $\Gamma_t(v)$, yielding $\mathcal{O}(\log |\Gamma_t(v)|)$ per hop and $\mathcal{O}(L \log |\Gamma_t(v)|)$ per walk. Cooperative dispatch (Section~\ref{sec:walk-scheduling}) reduces wall-clock time by amortizing the per-node work across co-located walks but does not change per-walk asymptotic cost.

\paragraph{Index construction and window management.}
Let $m = |W(t)|$ denote the number of edges in the active window. Each batch boundary triggers eviction and reconstruction of the dual-index. The reconstruction performs two GPU radix sorts (one for each view) and a constant number of linear-time prefix-sum and offset-generation passes. CUB's radix sort is linear in input size~\cite{cub}, so each runs in $\mathcal{O}(m)$ work, and the per-batch reconstruction cost is $\mathcal{O}(m)$. These costs are amortized across the $K$ walks generated from the rebuilt indices, giving a per-walk amortized index cost of $\mathcal{O}(m/K)$.

\paragraph{Memory usage.}
Memory is bounded by the active window. Each edge in $W(t)$ is stored once in the shared edge store, with auxiliary index arrays of total size linear in $m$. No memory is retained for evicted edges, and usage does not grow with input stream length.

\section{Evaluation}

\subsection{Experimental Setup}
\label{sec:setup}

\paragraph{Hardware and configuration.}
Experiments use compute nodes with 32 CPU cores, 196 GB of RAM, and an NVIDIA A40 GPU (46 GB VRAM, CUDA 12.6). GPU kernels use a block dimension of 256 (Section~\ref{sec:parameter-tuning}). Walk length is 80, walks per node is 10, temporal bias is exponential, and start bias is uniform unless stated otherwise. The default sampler is weight-based (Section~\ref{sec:temporal-bias}). Streaming experiments process temporally ordered batches under a sliding window of one-third the dataset's time span. Results average over five runs.

\paragraph{Baselines.}
We compare against \textbf{TEA+}~\cite{teaplus2024} and \textbf{TEA}~\cite{tea2023}, CPU temporal-walk engines using hybrid alias sampling; for temporal walks the two systems are identical, as evidenced by TEA+ reporting the same tables as TEA (TEA: Table~4, TEA+: Table~2). Several random walk engines target static or dynamic-structure graphs, including GraphWalker~\cite{graphwalker2020}, KnightKing~\cite{knightking2019}, FlashMob~\cite{flashmob2021}, ThunderRW~\cite{thunderrw2021}, FlowWalker~\cite{flowwalker2024}, and Bingo~\cite{bingo2025}; these systems do not ingest timestamps and cannot enforce temporal causality. We compare against two representatives in Section~\ref{sec:non-temporal-baseline-comparison}.

\subsection{Datasets}
\label{sec:datasets}

We report streaming-scale results on Alibaba, the largest publicly available temporal interaction dataset (81B edges over 14 days)~\cite{luo2022workload}. Detailed ablation, profiling, and sensitivity studies use TGBL-Coin, TGBL-Flight~\cite{tgb2023}, and Konect-Delicious~\cite{kunegis2013konect}, three of the largest temporal datasets in TGB and Konect. Profiling at Alibaba scale is impractical due to the cost of repeated full-stream replays. Table~\ref{tab:datasets} lists the datasets used in the main paper.

\begin{table}[h]
\centering
\caption{Datasets used in evaluation.}
\label{tab:datasets}
\begin{tabular}{lcc}
\toprule
\textbf{Dataset} & \textbf{Nodes} & \textbf{Edges} \\
\midrule
TGBL-Review~\cite{tgb2023} & 352K & 4.8M \\
TGBL-Coin~\cite{tgb2023} & 638.5K & 22.8M \\
Konect-Growth~\cite{kunegis2013konect} & 1.8M & 39M \\
TGBL-Comment~\cite{tgb2023} & 994.8K & 44.3M \\
TGBL-Flight~\cite{tgb2023} & 18K & 67M \\
Konect-Delicious~\cite{kunegis2013konect} & 33.7M & 301M \\
Alibaba Microservices~\cite{luo2022workload} & 68K & 81B \\
Hub-Synthetic & 24.6K & 27M \\
\bottomrule
\end{tabular}
\end{table}

\subsection{Streaming-Scale Evaluation}
\label{sec:streaming_evaluation}

\begin{figure}[h]
    \centering
    \begin{subfigure}[b]{0.49\linewidth}
        \centering
        \includegraphics[width=\linewidth]{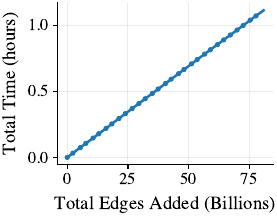}
        \caption{Cumulative edge ingestion time.}
    \end{subfigure}\hfill
    \begin{subfigure}[b]{0.49\linewidth}
        \centering
        \includegraphics[width=\linewidth]{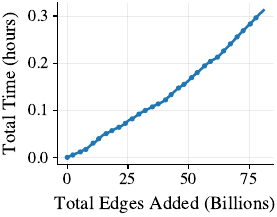}
        \caption{Cumulative walk sampling time.}
    \end{subfigure}
    \caption{Cumulative streaming performance on the Alibaba dataset over 81B edges.}
    \label{fig:alibaba_streaming_performance}
\end{figure}

We evaluate Tempest under sustained streaming on the Alibaba 2022 Microservice Logs~\cite{luo2022workload}: 81 billion interactions among 68,000 microservices over 14 days, partitioned into 6{,}720 chronological 3-minute batches of $\approx$12M edges each. Tempest uses a 60-minute sliding window, holding $\approx$120M active edges in steady state. After each batch, the system generates 20 temporal random walks of length 100 from every active source node ($\approx$219K walks per batch). High-frequency interaction data of this kind concentrates many events into each millisecond timestamp, making timestamp gaps approximately uniform; we therefore use the index-based sampler with the closed-form $\mathcal{O}(1)$ inverse-CDF (Section~\ref{sec:temporal-bias}), which is exact under uniform-gap conditions.

\paragraph{Results.}
Figure~\ref{fig:alibaba_streaming_performance} shows cumulative ingestion and walk-sampling time as the 14-day stream is replayed. Tempest sustains real-time processing throughout: per-batch ingest averages 596 ms and walk sampling averages 167 ms, against a 180-second batch arrival interval, leaving over $235\times$ headroom and no backlog. The full 81 billion edges are processed end-to-end in 1.42 hours of wall time (1.11 hours ingestion, 0.31 hours sampling). The ingestion curve is essentially linear, confirming that batch-bounded reconstruction cost (Section~\ref{sec:streaming-ingestion}) does not accumulate as the stream advances.

\subsection{Scaling Behavior}
\label{sec:scaling-behavior}

\begin{figure}[h]
    \centering
    \begin{subfigure}[b]{0.49\linewidth}
        \centering
        \includegraphics[width=\linewidth]{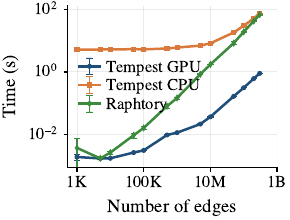}
        \caption{Edge ingestion.}
    \end{subfigure}\hfill
    \begin{subfigure}[b]{0.49\linewidth}
        \centering
        \includegraphics[width=\linewidth]{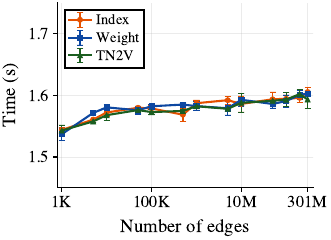}
        \caption{Walk sampling (10M walks).}
    \end{subfigure}
    \caption{Scaling on Konect-Delicious (1K to 301M edges).}
    \label{fig:scaling}
\end{figure}

We characterize how Tempest scales with active graph size on Konect-Delicious. To our knowledge, no random walk engine supports streaming ingestion as part of its pipeline; we therefore compare against Raphtory~\cite{raphtory2021}, an industrial temporal graph engine supporting incremental edge insertion at this scale, and against a CPU backend of Tempest itself. The CPU backend mirrors the GPU data-parallel primitives, pickers, and samplers using OpenMP and TBB; the hierarchical cooperative scheduler is GPU-only. Both backends share the same data structures and source tree, so CPU-vs-GPU comparisons isolate the GPU's parallelism advantage.

\paragraph{Ingestion.}
Each point on Figure~\ref{fig:scaling}a is a separate run that ingests the indicated number of edges from scratch and rebuilds the dual-index. At 301M edges, Tempest GPU completes ingestion in 0.9\,s, $76\times$ faster than Raphtory and $84\times$ faster than the same algorithm on CPU. Tempest CPU and Raphtory are within 10\% of each other across the range, which separates the GPU advantage from any algorithmic difference between Tempest and Raphtory.

\paragraph{Walk sampling.}
Each point on Figure~\ref{fig:scaling}b is a separate run that ingests the indicated number of edges and then generates 10M walks of length 100 with uniformly sampled start nodes. All three pickers stay essentially flat across edge counts: per-walk time varies by less than 5\% from 1K to 301M edges. The rejection-sampling formulation of Temporal Node2Vec (Section~\ref{sec:temporal-bias}) lets the second-order picker run through the same cooperative dispatch path as the first-order pickers, with per-hop cost dominated by walk length rather than neighborhood size. The GPU backend accelerates all three pickers by 2.5--2.9$\times$ over the CPU backend (CPU lines are omitted from the figure for brevity).

\subsection{Parameter Tuning}
\label{sec:parameter-tuning}

The cooperative scheduler exposes two tunable parameters: the CUDA block dimension and the solo--warp boundary $W_{\mathrm{warp}}$.

\begin{figure}[h]
    \centering
    \begin{subfigure}[b]{0.49\linewidth}
        \centering
        \includegraphics[width=\linewidth]{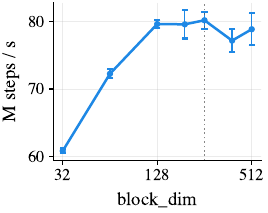}
        \caption{Walk throughput.}
    \end{subfigure}\hfill
    \begin{subfigure}[b]{0.49\linewidth}
        \centering
        \includegraphics[width=\linewidth]{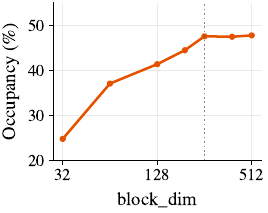}
        \caption{SM occupancy.}
    \end{subfigure}
    \caption{Block dimension sweep on TGBL-Coin.}
    \label{fig:block-dim-tuning}
\end{figure}

\paragraph{Block dimension.}
Figure~\ref{fig:block-dim-tuning} sweeps the block dimension on TGBL-Coin. Block dimension is determined by the kernel's GPU resource footprint, so a single representative dataset suffices. Throughput rises sharply through 128 and then flattens; values from 128 to 512 lie within run-to-run standard deviation of each other. SM occupancy continues to rise past 128 and saturates at 256. Going beyond 256 also requires opting into the architecture's higher dynamic shared-memory ceiling, which causes portability and occupancy issues. We default to 256, the largest block size that fits the static shared-memory envelope and captures both the throughput plateau and the occupancy saturation point.

\begin{figure}[h]
    \centering
    \includegraphics[width=0.75\linewidth]{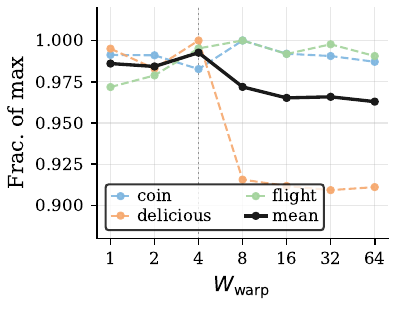}
    \caption{$W_{\mathrm{warp}}$ sweep across three datasets. Per-dataset throughput is normalized to its own maximum; the bold curve is the cross-dataset mean.}
    \label{fig:w-warp-tuning}
\end{figure}

\paragraph{Solo--warp boundary.}
Figure~\ref{fig:w-warp-tuning} sweeps $W_{\mathrm{warp}} \in \{1, 2, 4, 8, 16, 32, 64\}$ on TGBL-Coin, TGBL-Flight, and Konect-Delicious. Per-dataset throughput stays within 3\% of its maximum on TGBL-Coin and TGBL-Flight across the full range. Konect-Delicious peaks at $W_{\mathrm{warp}} = 4$ and drops by 9\% at $W_{\mathrm{warp}} \ge 8$. Its heavy-tailed degree distribution produces many medium-$W$ nodes that benefit from warp-coop amortization. At higher thresholds these nodes fall into solo dispatch, forfeiting the amortization. The cross-dataset mean peaks at $W_{\mathrm{warp}} = 4$, which we adopt as the default.

\subsection{Cooperative Scheduler Ablation}
\label{sec:scheduler-ablation}

\begin{table*}[h]
\centering
\small
\setlength{\tabcolsep}{4pt}
\caption{Cooperative scheduler ablation. Steps/sec averaged over five runs (standard deviations within 2\% across all entries); kernel time and launch count are per walk-generation call from nsys profiling.}
\label{tab:scheduler-ablation}
\begin{tabular}{l|ccc|ccc|ccc|ccc}
\toprule
& \multicolumn{3}{c|}{\textbf{TGBL-Coin}} & \multicolumn{3}{c|}{\textbf{Konect-Delicious}} & \multicolumn{3}{c|}{\textbf{tgbl-comment}} & \multicolumn{3}{c}{\textbf{Hub-Synthetic}} \\
\textbf{Variant} & M-steps/s & ms/call & launches & M-steps/s & ms/call & launches & M-steps/s & ms/call & launches & M-steps/s & ms/call & launches \\
\midrule
Full-Walk    &  67.8 & 2{,}723 &     8 & 101.4 & 457 &   8 & 69.5 & 2{,}225 & 8 & 117.8 & 6{,}396 &     8 \\
Coop-Global  &  72.0 & 2{,}294 & 2{,}456 & 106.7 & 365 & 283 & 86.1 & 1{,}311 & 2{,}456 & 141.5 & 4{,}492 & 2{,}880 \\
Coop         &  73.3 & 2{,}283 & 2{,}613 & 108.4 & 346 & 283 & 88.1 & 1{,}340 & 2{,}608 & 170.3 & 3{,}314 & 2{,}880 \\
\bottomrule
\end{tabular}
\end{table*}

We isolate the two mechanisms of the cooperative scheduler (Section~\ref{sec:walk-scheduling}) by comparing three execution variants on TGBL-Coin, Konect-Delicious, tgbl-comment, and Hub-Synthetic. Hub-Synthetic is a synthetic temporal graph (24,600 nodes, 27M edges) included to isolate the scheduler's contribution under workloads where benchmark datasets each fail one or more of the structural conditions the scheduler exploits.

\paragraph{Variants.}
\textbf{Full-Walk} is the one-thread-per-walk baseline of Section~\ref{sec:full-walk-baseline}: each thread carries one walk to completion with no shared state. \textbf{Coop-Global} enables the per-step regrouping and tier dispatch but disables the smem metadata; cooperative tiers read the per-node timestamp-group offsets from global memory on every hop. \textbf{Coop} is the full scheduler: per-step regrouping plus tier dispatch plus the smem panel.

\paragraph{Results.}

Table~\ref{tab:scheduler-ablation} shows that cooperative dispatch alone (Coop-Global vs.~Full-Walk) lifts throughput by up to 24\%, and adding smem-resident metadata (Coop vs.~Coop-Global) lifts it by up to a further 20\%, for an end-to-end gain of up to 45\%. The largest gains appear on workloads where walks concentrate on hub nodes through their full duration: tgbl-comment walks survive 37 hops on hub-dense thread structures, and Hub-Synthetic walks survive 97 by construction. Coin and Delicious distribute walks more broadly across the graph (Coin survives 34 hops with moderate hub skew; Delicious walks terminate at 5 hops), reducing the per-step convergence the scheduler amortizes.

The structural change is visible in the launch and per-call columns. Full-Walk performs each walk-generation call with eight monolithic kernel launches, running for hundreds of milliseconds to several seconds (2.7\,s on coin, 6.4\,s on Hub-Synthetic). Coop and Coop-Global decompose work into hundreds to thousands of tier-specific micro-kernels: $\sim$2{,}600 launches on coin and tgbl-comment (320$\times$ more), $\sim$2{,}880 on Hub-Synthetic (360$\times$ more), and $\sim$280 on delicious due to its short walks (35$\times$ more). Per-kernel time drops to tens or hundreds of microseconds, and wall-clock per call drops by up to 48\% despite the larger launch count. Smem-resident metadata accounts for the gain over Coop-Global by amortizing the per-node temporal-index access across co-located walks, reducing global-memory traffic in warp- and block-tier kernels. Beyond the throughput gain, the scheduler bounds per-launch cost regardless of walk-population skew, enabling scaling to mega-hub workloads.

\begin{table}[h]
\centering
\caption{Per-tier launch distribution (\%) on Coop.}
\label{tab:tier-distribution}
\begin{tabular}{lrrr}
\toprule
\textbf{Tier} & \textbf{Coin} & \textbf{Delicious} & \textbf{Comment} \\
\midrule
solo                  & 14.2 & 20.2 & 14.3 \\
warp $\times$ smem    & 14.4 & 20.5 & 14.5 \\
warp $\times$ global  & 14.4 &  0.0 & 14.5 \\
block $\times$ smem   & 14.2 & 16.9 & 13.4 \\
block $\times$ global & 14.2 &  0.0 & 14.3 \\
multi-block           & 28.7 & 42.5 & 29.0 \\
\bottomrule
\end{tabular}
\end{table}

\paragraph{Tier distribution.}
Table~\ref{tab:tier-distribution} shows that the dispatch plane is fully exercised across datasets. Mega-hub expansion accounts for 29--43\% of launches, validating the W-axis. Coin and tgbl-comment exercise the global tier where $G$ exceeds the smem cap, validating the G-axis; Delicious never reaches the global fallback because its $G$ distribution fits within the smem caps at every step.

\paragraph{Factor isolation.}
The scheduler's gain is the product of three workload properties:
\begin{itemize}[leftmargin=*]
    \item \textbf{Factor A — Walks-per-hub-per-step density ($W$):} selects the dispatch tier (solo, warp, or block) and determines the per-walk amortization factor, which differs significantly across tiers.
    \item \textbf{Factor B — Per-hub timestamp-group count ($G$) in the smem-fittable band:}  determines whether the smem panel preloads metadata to smem or falls back to the global tier.
    \item \textbf{Factor C — Walk persistence past the short-walk regime:} determines whether per-step savings amortize against fixed scheduler costs (alive-flag scan, sort, partition).
\end{itemize}

A controlled ablation on Hub-Synthetic disables each factor in isolation through four variants:
\begin{itemize}[leftmargin=*]
    \item \textbf{LowW} drops walks-per-node from 500 to 100, forcing walks below the block-tier threshold.
    \item \textbf{HighG} raises $G$ from 1{,}500 to 3{,}000, forcing the global-tier.
    \item \textbf{ShortWalk} truncates maximum walk length from 50 to 10, terminating walks before per-step amortization completes.
    \item \textbf{AllOff} disables all three factors simultaneously.
\end{itemize}

Each variant degrades performance relative to full Hub-Synthetic dataset: LowW reduces the gain from 45.8\% to 33.6\%, HighG to 21.8\%, ShortWalk to 35.6\%, and AllOff to 15.5\%. The drops are roughly additive, with AllOff bounded near the lower limit of scheduler effectiveness on this workload. Cross-dataset variation across the benchmark datasets reflects different combinations of these factors.

\begin{figure}[t]
    \centering
    \includegraphics[width=0.9\linewidth]{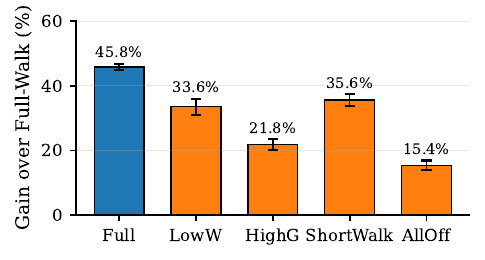}
    \caption{Factor-isolation ablation on Hub-Synthetic. Each variant disables one or more architectural factors that the cooperative scheduler exploits.}
    \label{fig:three-factor-ablation}
\end{figure}

\subsection{Ingestion Time Breakdown}
\label{sec:ingestion-breakdown}

We profile each batch with NVIDIA's NVTX to measure four stages: dual-index sort, cumulative-weight precomputation for bias sampling, host-to-device transfer, and pipeline overhead. Pipeline overhead captures the per-batch recurring framework costs (GPU memory allocation, kernel-launch queuing, and CPU-GPU coordination).

\begin{table*}[h]
\centering
\caption{Performance comparison on shared datasets. Runtimes in seconds; best values in bold.}
\label{tab:comparison_with_tea}
\begin{tabular}{lcccccc}
\toprule
& \multicolumn{2}{c}{\textbf{Exponential}} & \multicolumn{2}{c}{\textbf{Linear}} & \multicolumn{2}{c}{\textbf{Temporal Node2Vec}} \\
\cmidrule(lr){2-3} \cmidrule(lr){4-5} \cmidrule(lr){6-7}
\textbf{Dataset} & \textbf{TEA+} & \textbf{Tempest} & \textbf{TEA+} & \textbf{Tempest} & \textbf{TEA+} & \textbf{Tempest} \\
\midrule
growth    &  2.93 & \textbf{0.50} ($5.8\times$) &  0.56 & \textbf{0.49} &  3.52 & \textbf{0.51} ($6.9\times$) \\
delicious & 38.84 & \textbf{8.43} ($4.6\times$) &  \textbf{7.98} & 8.36 & 59.82 & \textbf{9.64 ($6.2\times$)} \\
\bottomrule
\end{tabular}
\end{table*}

\begin{table}[h]
\centering
\caption{Per-batch ingestion time breakdown (\% of total).}
\label{tab:ingestion-breakdown}
\begin{tabular}{lrrr}
\toprule
\textbf{Stage} & \textbf{Coin} & \textbf{Delicious} & \textbf{Comment} \\
\midrule
sort               &  6.3 &  8.9 & 11.7 \\
weight             & 10.8 & 26.4 &  9.5 \\
H2D                & 26.4 & 23.9 & 27.5 \\
pipeline overhead  & 56.5 & 40.9 & 51.3 \\
\midrule
total (ms)         & 210  & 226  & 211 \\
\bottomrule
\end{tabular}
\end{table}

\paragraph{Results.}
Table~\ref{tab:ingestion-breakdown} shows the breakdown averaged over a full streaming run. Sort accounts for 6--12\%, validating the linear-time CUB radix sort claimed in Section~\ref{sec:complexity}. Weight precomputation scales with node count: tgbl-comment (995K nodes) sees 9.5\%, Delicious (34M nodes) sees 26.4\%. Host to device transfer (H2D) is roughly proportional to batch size. Pipeline overhead dominates because each batch accumulates many small framework costs (memory allocation, kernel launches, CPU-GPU coordination) across the ingestion pipeline. Each component is bounded by the active window $|W(t)|$ and does not accumulate as the stream advances.

\subsection{Comparison with TEA+/TEA}
\label{sec:tea-comparison}

TEA+~\cite{teaplus2024} and its predecessor TEA~\cite{tea2023} are CPU-based engines for causality-preserving temporal random walks, the closest comparable baselines to Tempest. We compare architectural design and end-to-end performance on shared workloads.

\paragraph{Architecture.}
TEA+ operates in bulk mode and requires the full temporal graph in memory before walk generation; it does not support sliding-window eviction. Its sampler relies on hierarchical alias tables and hash-indexed structures; per-vertex alias trunks consume substantial auxiliary memory and require updates across multiple distributed structures when edges are evicted. Tempest stores edges once and exposes two logical orderings (Section~\ref{sec:dual-index}), supporting global start-edge selection and window eviction over the same physical layout, with hop-by-hop progression served from a node-local timestamp ordering.

\paragraph{Source code and reproducibility.}
TEA+'s source code is not publicly available and could not be obtained from the authors. Faithful reimplementation without access to internal design details would be unlikely to yield a fair comparison. Following the standard practice for evaluating closed-source systems~\cite{fitx2024,ffmalloc2021,wukong2016}, we report TEA+'s performance numbers as published.

\paragraph{Datasets.}
TEA+ reports four KONECT~\cite{kunegis2013konect} datasets. We restrict our comparison to growth and delicious, whose edge counts and temporal attributes match the public KONECT release. The reported edge count for edit differs from the count listed on KONECT, and twitter is listed on KONECT without timestamps. We therefore exclude both from the comparison.

\paragraph{Performance.}

Table~\ref{tab:comparison_with_tea} reports walk generation runtimes under the configuration TEA+ uses (1 walk per node, walk length 80). Tempest runs in bulk mode for parity, while its primary operating regime is streaming ingestion in smaller batches under bounded memory (Section~\ref{sec:streaming_evaluation}). Under exponential bias, the configuration TEA identifies as the most relevant for temporal walks~\cite{tea2023}, Tempest is $4.6$--$5.8\times$ faster. Under linear bias the two systems are comparable. For Temporal Node2Vec, Tempest is $6$--$7\times$ faster on both datasets. Due to differing execution models and hardware, these results are indicative rather than definitive.

\subsection{Window Duration Sensitivity}
\label{sec:window-size-sensitivity-analysis}

The temporal window duration $\Delta$ trades historical context against per-batch cost. We sweep $\Delta$ from 1 to 10 batches on TGBL-Coin, TGBL-Review, and TGBL-Flight, with batch duration fixed at $T/100$ where $T$ is each dataset's time span. Walk length, walks per node, and bias are held constant.

For downstream evaluation, we split each dataset chronologically into 70/15/15 train/validation/test partitions. Walks generated after each training batch update node embeddings via incremental skip-gram~\cite{word2vec2013} training. Link prediction is supervised against negative edges constructed by replacing each positive edge's target with a non-co-occurring node. We report mean and standard deviation of test AUC across five trials.

\begin{figure}[h]
    \centering
    \begin{subfigure}[b]{0.5\linewidth}
        \centering
        \includegraphics[width=\linewidth]{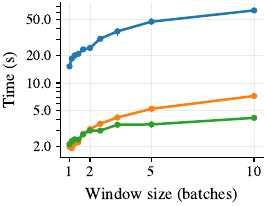}
        \caption{Walk sampling.}
    \end{subfigure}\hfill
    \begin{subfigure}[b]{0.5\linewidth}
        \centering
        \includegraphics[width=\linewidth]{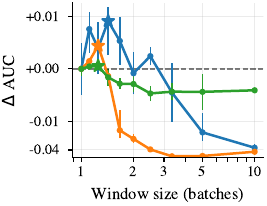}
        \caption{Link prediction $\Delta$AUC.}
    \end{subfigure}
    \vspace{0.5ex}
    \centering
    \includegraphics{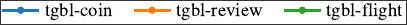}
    \caption{Window sensitivity on TGB datasets. (b) reports $\Delta$AUC relative to the smallest window (1 batch); baseline AUCs are 0.861 (coin), 0.535 (review), 0.971 (flight). Asterisks mark per-dataset optima.}
    \label{fig:window_sensitivity_analysis}
\end{figure}

\begin{table*}[h]
\centering
\caption{Comparison with non-temporal walk engines. Throughput is mean $\pm$ std of 10 runs. Best values in bold.}
\label{tab:non-temporal-comparison}
\begin{tabular}{@{}l r ccc ccc ccc@{}}
\toprule
& \textbf{Avg.} & \multicolumn{3}{c}{\textbf{Throughput (M steps/s)}} & \multicolumn{3}{c}{\textbf{Valid Hops (\%)}} & \multicolumn{3}{c}{\textbf{Valid Walks (\%)}} \\
\cmidrule(lr){3-5} \cmidrule(lr){6-8} \cmidrule(lr){9-11}
\textbf{Dataset} & \textbf{Degree} & \textbf{Tempest} & \textbf{Flow.} & \textbf{Thun.} & \textbf{Tempest} & \textbf{Flow.} & \textbf{Thun.} & \textbf{Tempest} & \textbf{Flow.} & \textbf{Thun.} \\
\midrule
Growth      & 21.4   & $\mathbf{30.6 \pm 0.6}$  & $28.8 \pm 0.0$  & $7.3 \pm 1.3$  & $\mathbf{100}$ & 0.0 & 1.1 & $\mathbf{100}$ & 0.0 & 0.0 \\
Delicious   & 8.9    & $\mathbf{66.7 \pm 1.2}$  & $1.3 \pm 0.0$   & $11.0 \pm 1.2$ & $\mathbf{100}$ & 0.0 & 0.1 & $\mathbf{100}$ & 0.0 & 0.0 \\
TGBL-Coin   & 35.7   & $\mathbf{63.8 \pm 1.2}$  & $10.0 \pm 0.0$  & $13.9 \pm 0.1$ & $\mathbf{100}$ & 0.0 & 0.9 & $\mathbf{100}$ & 0.0 & 0.0 \\
TGBL-Flight & 3702.2 & $\mathbf{112.3 \pm 1.5}$ & $97.3 \pm 0.1$  & $16.6 \pm 0.3$ & $\mathbf{100}$ & 0.0 & 1.2 & $\mathbf{100}$ & 0.0 & 0.0 \\
\bottomrule
\end{tabular}
\end{table*}

\paragraph{Results.}
Figure~\ref{fig:window_sensitivity_analysis} shows walk-sampling latency rising monotonically with $\Delta$ as the active edge set grows. Downstream link prediction peaks at $\Delta = 1$--$2$ batches; beyond that, AUC plateaus or declines. Larger windows resample stale interactions across batches, diluting the contribution of recent edges. Recent context alone is sufficient for accurate prediction when the underlying model is updated regularly; increasing $\Delta$ beyond this regime spends GPU memory and sampling cost without improving prediction quality.

\subsection{Comparison with Non-Temporal Baselines}
\label{sec:non-temporal-baseline-comparison}

GPU and CPU random walk systems for static graphs operate on time-agnostic abstractions and do not enforce temporal causality. For consistency with prior comparisons in this space \cite{teaplus2024, tea2023}, we evaluate Tempest against two recent, high-performance static frameworks for which open-source artifacts are available: the GPU-based FlowWalker~\cite{flowwalker2024} and the CPU-based ThunderRW~\cite{thunderrw2021}. We ran FlowWalker on the same NVIDIA A40 used for Tempest and ThunderRW on a 20-core CPU server with 192~GB of RAM.

\paragraph{Setup.}
We evaluate on four temporal datasets at varying scales. For FlowWalker and ThunderRW, timestamps were discarded to accommodate their static graph abstraction. With each system we generate 10 million walks of length 80. Temporal causality was independently validated by post-processing each walk using a greedy, earliest-feasible timestamp assignment rule. Walks violating strict timestamp monotonicity were marked as invalid.

\paragraph{Results.}
Table~\ref{tab:non-temporal-comparison} reports throughput alongside hop- and walk-level temporal validity. FlowWalker and ThunderRW produce 0\% valid walks across every dataset: although a small fraction of individual hops happen to satisfy temporal monotonicity by chance (up to 1.2\%), every walk contains at least one violation. This is structural, not a tuning issue: static engines have no notion of temporal causality and cannot preserve it. Once the validity gap is established, the throughput numbers quantify the cost of enforcing causality. Tempest is fastest on every dataset, with the largest margins on Delicious ($51\times$ over FlowWalker) and TGBL-Coin ($6.4\times$). Throughput scales with the average static degree across datasets.

\subsection{Memory Usage Analysis}
\label{sec:memory-analysis}

We evaluate Tempest's memory footprint under two regimes: bulk mode with varying edge count, and streaming mode with a sliding window where each ingested batch evicts the previous one.

\begin{figure}[h]
    \centering
    \begin{subfigure}[b]{0.5\linewidth}
        \centering
        \includegraphics[width=\linewidth]{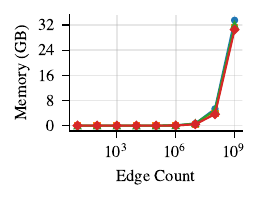}
        \caption{}
    \end{subfigure}\hfill
    \begin{subfigure}[b]{0.5\linewidth}
        \centering
        \includegraphics[width=\linewidth]{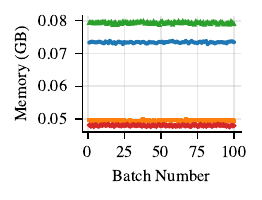}
        \caption{}
    \end{subfigure}
    \vspace{1ex}
    \includegraphics[width=\linewidth]{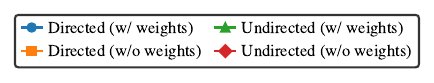}
    \caption{Memory usage. (a) Edge scaling in bulk mode (nodes = 10K, timestamps = 10K). (b) Streaming with a constant window over 100 batches of 10M edges each.}
    \label{fig:memory_usage}
\end{figure}

\paragraph{Results.}
Figure~\ref{fig:memory_usage}a shows linear memory growth with edge count, matching Tempest's flat layout: 1 billion edges fit in approximately 35~GB, well within commodity datacenter GPUs like the NVIDIA A40 (46~GB VRAM). Figure~\ref{fig:memory_usage}b shows memory remaining flat across 100 streaming batches, confirming the bounded-memory claim from Section~\ref{sec:complexity}: memory does not grow with stream length, only with active-window size. Weight-based sampling consumes slightly more memory than index-based due to the cumulative-weight arrays, and trends hold for directed and undirected modes.

\section{Conclusion}
Tempest is a GPU-accelerated engine for streaming temporal random walks combining a dual-index edge store with a hierarchical cooperative scheduler that dispatches walks at thread, warp, or block granularity. Sustained processing of 81 billion edges over a 14-day stream demonstrates bounded-memory streaming temporal walks at industrial scale, with $76\times$ ingestion speedup over the closest temporal graph engine and causal correctness against non-temporal engines that produce $0\%$ valid walks. The cooperative scheduler delivers measurable end-to-end gains on hub-skewed temporal graphs by bounding per-launch work cost. Tempest opens a path to walk-native temporal embedding methods that consume causal walks directly, which we are pursuing in ongoing work.

\FloatBarrier

\begin{acks}
Md Ashfaq Salehin is supported by a Sussex AI PhD Studentship. The authors thank Dr James Knight (University of Sussex) for helpful suggestions on CUDA best practices.
\end{acks}

\bibliographystyle{ACM-Reference-Format}
\bibliography{references}

\end{document}